# Generalized Local Fractional Taylor's Formula with Local Fractional Derivative

## Xiao-Jun Yang

Department of Mathematics and Mechanics, China University of Mining and Technology, Xuzhou Campus, Xuzhou, Jiangsu, 221008, P. R. China

Email: dyangxiaojun@163.com

**Abstract** –In the present paper, a generalized local Taylor formula with the local fractional derivatives (LFDs) is proposed based on the local fractional calculus (LFC). From the fractal geometry point of view, the theory of local fractional integrals and derivatives has been dealt with fractal and continuously non-differentiable functions, and has been successfully applied in engineering problems. It points out the proof of the generalized local fractional Taylor formula, and is devoted to the applications of the generalized local fractional Taylor formula to the generalized local fractional series and the approximation of functions. Finally, it is shown that local fractional Taylor series of the Mittag-Leffler type function is discussed.

## 1. Introduction

The local fractional Taylor formula has been generalized by many authors. Kolwankar and Gangal had already written a classically formal version of the local fractional Taylor series [1, 2]

$$f(x) = \sum_{i=0}^{n} \frac{f^{(n)}(y)}{\Gamma(1+n)}(x-y)^n + \frac{D^\alpha f(y)}{\Gamma(1+\alpha)}(x-y)^\alpha + R_\alpha(x,y) \quad (1.1)$$

where $D^\alpha f(y)$ is the Kolwankar and Gangal local fractional derivatives, denoted by

$$D^\alpha f(y) = \lim_{x \to y} \frac{d^\alpha [f(x)-f(y)]}{[d(x-y)]^\alpha} \quad (1.2)$$

and its reminder is

$$R_\alpha(x,y) = \frac{1}{\Gamma(1+\alpha)} \int_0^{x-y} \frac{dF(y,t,\alpha,n)}{dt}(x-y-t)^\alpha dt \quad (1.3)$$

where $F(y,x-y,\alpha) = \frac{d^\alpha(f(x)-f(y))}{[d(x-y)]^\alpha}$.

On the other hand, Adda and Cresson obtained the following relation [3]

$$f(x) = f(y) + \frac{d_\sigma^\alpha f(y)}{\Gamma(1+\alpha)}[\sigma(x-y)]^\sigma + R_\sigma(x,y) \quad (1.4)$$

with

$$R_\sigma(x,y) = \sigma \frac{1}{\Gamma(1+\alpha)} \int_0^{x-y} \frac{dF_\sigma(y,\sigma t,\alpha)}{dt}(\sigma(x-y-t))^\alpha dt$$

and

$$\lim_{x \to y^\sigma} \frac{R_\sigma(x,y)}{(\sigma(x-y))^\alpha} = 0,$$

where

$$F_\sigma(y,\sigma(x-y),\alpha) = D_{y,-\sigma}^\alpha[\sigma(f-f(y))](x)$$

and Adda and Cresson's local fractional derivative is denoted by

$$d_\sigma^\alpha f(y) = \lim_{x \to y^\sigma} D_{y,-\sigma}^\alpha[\sigma(f-f(y))](x). \quad (1.5)$$

Recently, Yang and Gao proposed the generalized local fractional Taylor series to study the Newton iteration method and introduced the following generalized local fractional Taylor series [7]

$$f(x) = \sum_{k=0}^{\infty} \frac{f^{(k\alpha)}(x_0)}{\Gamma(1+k\alpha)}(x-x_0)^{k\alpha} \quad (1.6)$$

with $a < x_0 < \xi < x < b$, $\forall x \in (a,b)$, and Gao-Yang-Kang local fractional derivative is denoted by [4-8]

$$f^{(\alpha)}(x_0) = \frac{d^\alpha f(x)}{dx^\alpha}\bigg|_{x=x_0} = \lim_{x \to x_0} \frac{\Delta^\alpha(f(x)-f(x_0))}{(x-x_0)^\alpha}, \quad (1.7)$$

with $\Delta^\alpha(f(x)-f(x_0)) \cong \Gamma(1+\alpha)\Delta(f(x)-f(x_0))$.

Successively, the sequential local fractional derivatives is denoted by

$$f^{(k\alpha)}(x) = \overbrace{D_x^{(\alpha)}...D_x^{(\alpha)}}^{k\ times} f(x) \quad (1.8)$$

If there exists the relation

$$|f(x)-f(x_0)|<\varepsilon^\alpha \quad (1.9)$$

with $|x-x_0|<\delta$, for $\varepsilon,\delta>0$ and $\varepsilon,\delta\in\mathbb{R}$.

Then $f(x)$ is called local fractional continuous on the interval $(a,b)$, denoted by

$$f(x)\in C_\alpha(a,b). \quad (1.10)$$

and sequential local fractional continuity is denoted by

$$C_\alpha^k(a,b) \quad (1.11)$$

or

$$f(x)\in C_\alpha^k(a,b).$$

However, the proof of the generalized local fractional Taylor series is not given. As a pursuit of the work we give some results for generalized local fractional Taylor formula by using the generalized mean value theorem for local fractional integrals and prove it.

This paper is organized as follows: In section 2, a brief introduction of local fractional derivative and integral are given. The generalized local Taylor's formula with local fractional derivative is investigated in section 3. Section 4 is devoted to the applications of the generalized local fractional Taylor formula to generalized local fractional series and approximation of functions. Conclusions are in section 5.

## 2. Preliminaries

**Definition 1**

Let $f(x)$ is local fractional continuous on the interval $[a,b]$. Local fractional integral of $f(x)$ of order $\alpha$ in the interval $[a,b]$ is defined [4, 6-7]

$$\begin{aligned}{}_aI_b^{(\alpha)}f(x) &= \frac{1}{\Gamma(1+\alpha)}\int_a^b f(t)(dt)^\alpha \\ &= \frac{1}{\Gamma(1+\alpha)}\lim_{\Delta t\to 0}\sum_{j=0}^{j=N-1}f(t_j)(\Delta t_j)^\alpha\end{aligned} \quad (2.1)$$

where $\Delta t_j = t_{j+1}-t_j$, $\Delta t=\max\{\Delta t_1,\Delta t_2,\Delta t_j,...\}$, and $[t_j,t_{j+1}]$ for $j=0,...,N-1$, $t_0=a, t_N=b$, is a partition of the interval $[a,b]$.

Here, it follows that

$${}_aI_a^{(\alpha)}f(x)=0 \text{ if } a=b; \quad (2.2)$$

$${}_aI_b^{(\alpha)}f(x)=-{}_bI_a^{(\alpha)}f(x) \text{ if } a<b; \quad (2.3)$$

and ${}_aI_a^{(0)}f(x)=f(x). \quad (2.4)$

Properties of the operator can be found in [6]. We only need here the following results:

For any $f(x)\in C_\alpha(a,b)$, $0<\alpha\le 1$, we have

$${}_{x_0}I_x^{(k\alpha)}f(x)=\overbrace{{}_{x_0}I_x^{(\alpha)}...{}_{x_0}I_x^{(\alpha)}}^{k\text{ times}}f(x); \quad (2.5)$$

$${}_{x_0}I_x^{(k\alpha)}x^{k\alpha}=\frac{\Gamma(1+k\alpha)}{\Gamma(1+(k+1)\alpha)}x^{(k+1)\alpha}. \quad (2.6)$$

For $0<\alpha\le 1$, $f^{(k\alpha)}(x)\in C_\alpha^k(a,b)$, then we have

$$\left({}_{x_0}I_x^{(k\alpha)}f(x)\right)^{(k\alpha)}=f(x), \quad (2.7)$$

where

$${}_{x_0}I_x^{(k\alpha)}f(x)=\overbrace{{}_{x_0}I_x^{(\alpha)}...{}_{x_0}I_x^{(\alpha)}}^{k\text{ times}}f(x)$$

and

$$f^{(k\alpha)}(x)=\overbrace{D_x^{(\alpha)}...D_x^{(\alpha)}}^{k\text{ times}}f(x).$$

For $f(x)=g^{(\alpha)}(x)\in C_\alpha[a,b]$, then we have [6]

$${}_aI_b^{(\alpha)}f(x)=g(b)-g(a). \quad (2.8)$$

**Theorem 1 (Mean value theorem for local fractional integrals)**

Suppose that $f(x)\in C_\alpha[a,b]$, we have [6]

$${}_aI_b^{(\alpha)}f(x)=f(\xi)\frac{(b-a)^\alpha}{\Gamma(\alpha+1)}, \quad a<\xi<b. \quad (2.9)$$

**Theorem 2**

Suppose that $f^{(k\alpha)}(x), f^{((k+1)\alpha)}(x)\in C_\alpha(a,b)$, for $0<\alpha\le 1$, then we have

$$\begin{aligned}&{}_{x_0}I_x^{(k\alpha)}[f^{(k\alpha)}(x)]-{}_{x_0}I_x^{((k+1)\alpha)}[f^{((k+1)\alpha)}(x)]\\&=f^{(k\alpha)}(\xi)\frac{(x-x_0)^{k\alpha}}{\Gamma(k\alpha+1)}\end{aligned}, \quad (2.10)$$

with $a<x_0<\xi<x<b$, where

$${}_{x_0}I_x^{((k+1)\alpha)}f(x)=\overbrace{{}_{x_0}I_x^{(\alpha)}...{}_{x_0}I_x^{(\alpha)}}^{k+1\text{ times}}f(x)$$

and

$$f^{((k+1)\alpha)}(x)=\overbrace{D_x^{(\alpha)}...D_x^{(\alpha)}}^{k+1\text{ times}}f(x).$$

*Proof.* From (2.5) and (2.9), we have

$$\begin{aligned}&{}_{x_0}I_x^{((k+1)\alpha)}[f^{((n+1)\alpha)}(x)]\\&={}_{x_0}I_x^{(k\alpha)}\left[\frac{1}{\Gamma(1+\alpha)}\int_{x_0}^x f^{((n+1)\alpha)}(x)(dt)^\alpha\right] \quad (2.11)\\&={}_{x_0}I_x^{(k\alpha)}\left(f^{(k\alpha)}(x)-f^{(k\alpha)}(\xi)\right) \quad (2.12)\end{aligned}$$

$$= {}_{x_0}I_x^{(k\alpha)} f^{(k\alpha)}(x) - {}_{x_0}I_x^{(k\alpha)} f^{(k\alpha)}(\xi). \quad (2.13)$$

Successively, it follows from (2.13) that

$$\begin{aligned} &{}_{x_0}I_x^{(k\alpha)} f^{(k\alpha)}(\xi) \\ &= f^{(k\alpha)}(\xi) \, {}_{x_0}I_x^{(k\alpha)} 1 \\ &= f^{(k\alpha)}(\xi) \, {}_{x_0}I_x^{((k-1)\alpha)} \left[ \frac{1}{\Gamma(1+\alpha)}(x-x_0)^\alpha \right] \\ &= f^{(k\alpha)}(\xi) \, {}_{x_0}I_x^{((k-2)\alpha)} \left[ \frac{\Gamma(1+\alpha)}{\Gamma(1+2\alpha)} \bullet \frac{1}{\Gamma(1+\alpha)}(x-x_0)^{2\alpha} \right] \\ &= f^{(k\alpha)}(\xi) \frac{(x-x_0)^{k\alpha}}{\Gamma(k\alpha+1)} \end{aligned} \quad (2.14)$$

Hence we have the result.

Remark. When $k=0$, considering the formula

$${}_{x_0}I_x^0[f^{(0)}(x)] - f(x) + f(x_0) = f(x_0),$$

we have

$${}_aI_x^0[f^{(0)}(x)] = f(x).$$

**Theorem 3 (Generalized mean value theorem for local fractional integrals)**

Suppose that $f(x) \in C_\alpha[a,b], f^{(\alpha)}(x) \in C(a,b)$, we have

$$f(x) - f(x_0) = f^{(\alpha)}(\xi)\frac{(x-x_0)^\alpha}{\Gamma(\alpha+1)}, \quad (2.15)$$

$a < x_0 < \xi < x < b$.

*Proof.* Taking $k=1$ in (2.10), we deduce to the result.

## 3. Generalized Local Fractional Taylor's Formula

In this section we will introduce a new generalization of local fractional Taylor formula that involving local fractional derivatives. We will begin with the mean value theorem for local fractional integrals.

**Theorem 4 (Generalized local fractional Taylor formula)**

Suppose that $f^{((k+1)\alpha)}(x) \in C_\alpha(a,b)$, for $k=0,1,...,n$ and $0<\alpha\leq 1$, then we have

$$f(x) = \sum_{k=0}^{n} \frac{f^{(k\alpha)}(x_0)}{\Gamma(1+k\alpha)}(x-x_0)^{k\alpha} + \frac{f^{((n+1)\alpha)}(\xi)}{\Gamma(1+(n+1)\alpha)}(x-x_0)^{(n+1)\alpha} \quad (3.1)$$

with $a<x_0<\xi<x<b$, $\forall x\in(a,b)$, where

$$f^{((k+1)\alpha)}(x) = \overbrace{D_x^{(\alpha)}...D_x^{(\alpha)}}^{k+1 \text{ times}} f(x).$$

*Proof.* Form (2.10), we have

$${}_{x_0}I_x^{(k\alpha)}[f^{(k\alpha)}(x)] - {}_{x_0}I_x^{((k+1)\alpha)}[f^{((k+1)\alpha)}(x)] = f^{(k\alpha)}(a)\frac{(x-x_0)^{k\alpha}}{\Gamma(k\alpha+1)}. \quad (3.2)$$

Successively, it follows from (3.2) that

$$\sum_{k=0}^{n} \left( {}_{x_0}I_x^{(k\alpha)}[f^{(k\alpha)}(x)] - {}_{x_0}I_x^{((k+1)\alpha)}[f^{((k+1)\alpha)}(x)] \right) \quad (3.3)$$

$$= f(x) - {}_{x_0}I_x^{((n+1)\alpha)}[f^{((n+1)\alpha)}(x)]$$

$$= \sum_{k=0}^{n} f^{(k\alpha)}(x_0) \frac{(x-x_0)^{k\alpha}}{\Gamma(k\alpha+1)}. \quad (3.4)$$

Applying (2.9) and (3.4), we have

$$\begin{aligned} &{}_{x_0}I_x^{((n+1)\alpha)}\left[ f^{((n+1)\alpha)}(x) \right] \\ &= \frac{1}{\Gamma(1+\alpha)} \int_{x_0}^{x} {}_aI_{x_0}^{(n\alpha)} f^{((n+1)\alpha)}(x)(dt)^\alpha \end{aligned} \quad (3.5)$$

$$= \frac{{}_aI_{x_0}^{(n\alpha)}\left[ f^{((n+1)\alpha)}(\xi)(x-x_0)^\alpha \right]}{\Gamma(1+\alpha)} \quad (3.6)$$

$$= f^{((n+1)\alpha)}(\xi) \frac{{}_aI_{x_0}^{(n\alpha)}(x-x_0)^\alpha}{\Gamma(1+\alpha)} \quad (3.7)$$

$$= \frac{f^{((n+1)\alpha)}(\xi)(x-x_0)^{(n+1)\alpha}}{\Gamma(1+(n+1)\alpha)} \quad (3.8)$$

with $a<x_0<\xi<x<b$, $\forall x\in(a,b)$.

Combing the formulas (3.4) and (3.8) in (3.2), we have the result.

**Theorem 5**

Suppose that $f^{((k+1)\alpha)}(x) \in C_\alpha(a,b)$, for $k=0,1,...,n$ and $0<\alpha\leq 1$, then we have

$$f(x) = \sum_{k=0}^{n} \frac{f^{(k\alpha)}(0)}{\Gamma(1+k\alpha)}x^{k\alpha} + \frac{f^{((n+1)\alpha)}(\theta x) x^{(n+1)\alpha}}{\Gamma(1+(n+1)\alpha)} \quad (3.9)$$

with $0<\theta<1$, $\forall x\in(a,b)$, where

$$f^{((k+1)\alpha)}(x) = \overbrace{D_x^{(\alpha)}...D_x^{(\alpha)}}^{k+1 \text{ times}} f(x).$$

*Proof.* Applying (3.1), for $a < x_0 < \xi < x < b$ and $x_0 = 0$, we have that

$$f(x) = \sum_{k=0}^{n} \frac{f^{(k\alpha)}(0)}{\Gamma(1+k\alpha)} x^{k\alpha} + \frac{f^{((n+1)\alpha)}(\xi) x^{(n+1)\alpha}}{\Gamma(1+(n+1)\alpha)}. \quad (3.10)$$

If $\xi = \theta x$, then we have

$$\frac{f^{((n+1)\alpha)}(\xi) x^{(n+1)\alpha}}{\Gamma(1+(n+1)\alpha)} = \frac{f^{((n+1)\alpha)}(\theta x) x^{(n+1)\alpha}}{\Gamma(1+(n+1)\alpha)} \quad (3.11)$$

with $0 < \theta < 1$.
Hence, the proof of the theorem is completed.

## 4. Applications: The Generalized Local Fractional Series and Approximation of Functions

**Theorem 6 (Generalized local fractional Taylor series)**

Suppose that $f^{((k+1)\alpha)}(x) \in C_\alpha(a,b)$, for $k = 0, 1, ..., n$ and $0 < \alpha \leq 1$, then we have

$$f(x) = \sum_{k=0}^{\infty} \frac{f^{(k\alpha)}(x_0)}{\Gamma(1+k\alpha)} (x-x_0)^{k\alpha} \quad (4.1)$$

with $a < x_0 < x < b$, $\forall x \in (a,b)$, where

$$f^{((k+1)\alpha)}(x) = \overbrace{D_x^{(\alpha)} ... D_x^{(\alpha)}}^{k+1 \text{ times}} f(x).$$

*Proof.* From (3.1), taking the reminder

$$R_n = \frac{f^{((n+1)\alpha)}(\xi) x^{(n+1)\alpha}}{\Gamma(1+(n+1)\alpha)} \quad (4.2)$$

as $n \to \infty$, we have the following relation

$$\lim_{n \to \infty} R_n = \frac{f^{((n+1)\alpha)}(\xi) x^{(n+1)\alpha}}{\Gamma(1+(n+1)\alpha)} = 0. \quad (4.3)$$

That is to say,

$$f(x) = \sum_{k=0}^{\infty} \frac{f^{(k\alpha)}(x_0)}{\Gamma(1+k\alpha)} (x-x_0)^{k\alpha}. \quad (4.4)$$

Therefore the theorem is proved.

**Theorem 7 (Generalized local fractional Mc-Laurin's series)**

Suppose that $f^{((k+1)\alpha)}(x) \in C_\alpha(a,b)$, for $k = 0, 1, ..., n$ and $0 < \alpha \leq 1$, then we have

$$f(x) = \sum_{k=0}^{\infty} \frac{f^{(k\alpha)}(0)}{\Gamma(1+k\alpha)} x^{k\alpha} \quad (4.5)$$

with $a < 0 < x < b$, $\forall x \in (a,b)$, where

$$f^{((k+1)\alpha)}(x) = \overbrace{D_x^{(\alpha)} ... D_x^{(\alpha)}}^{k+1 \text{ times}} f(x).$$

*Proof.* Taking $x_0 = 0$ in (4.1), we obtain the result.

**Theorem 8 (Theorem for approximation of functions)**

Suppose that $f^{((k+1)\alpha)}(x) \in C_\alpha(a,b)$, for $k = 0, 1, ..., n$ and $0 < \alpha \leq 1$, then we have

$$f(x) \cong \sum_{k=0}^{n=N} \frac{f^{(k\alpha)}(x_0)}{\Gamma(1+k\alpha)} (x-x_0)^{k\alpha} \quad (4.6)$$

with $a < x_0 < x < b$, $\forall x \in (a,b)$, where

$$f^{((k+1)\alpha)}(x) = \overbrace{D_x^{(\alpha)} ... D_x^{(\alpha)}}^{k+1 \text{ times}} f(x).$$

Furthermore, the error term $R_n^N$ has the form

$$R_n^N = \frac{f^{((N+1)\alpha)}(\xi) x^{(N+1)\alpha}}{\Gamma(1+(N+1)\alpha)}. \quad (4.7)$$

*Proof.* The proof follows directly form (3.1).

*Example*

The Mittag-Leffler function [8] with fractal dimension $\alpha$ is defined as

$$E_\alpha(x^\alpha) = \sum_{k=0}^{\infty} \frac{x^{\alpha k}}{\Gamma(1+k\alpha)}. \quad (4.8)$$

There exists a polynomial

$$E_\alpha(x^\alpha) \cong 1 + \frac{x^\alpha}{\Gamma(1+\alpha)} + \frac{x^{2\alpha}}{\Gamma(1+2\alpha)} + ... + \frac{x^{N\alpha}}{\Gamma(1+N\alpha)},$$

$N \in \mathbb{N}$.

## 5. Conclusions

This paper has pointed out the generalized local fractional Taylor formula with local fractional derivative. As well, we discussed local fractional Taylor' series with local fractional derivative. The generalized local fractional Taylor series seems to look like fractional Taylor's series with modified Riemann - Liouville derivative in the form [9]. However, the derivative of the former is described by local fractional derivative, the later is modified Riemann-Liouville derivative. The differences of them were discussed in [7, 9]. Hence, when we make use of the generalized local fractional Taylor formula with local fractional derivative, it is important to defer from them. For more details of the theory and applications of local fractional calculus, see [10-17].